\documentclass[transaction,10pt]{IEEEtran}
\usepackage{graphicx}
\usepackage{eqnarray,amsmath}
\usepackage{times}
\usepackage{gfcoeff}
\usepackage{setspace}
\usepackage{epstopdf}

\begin{document}
\title{Ultra-Low Energy and High Speed LIF Neuron using Silicon Bipolar Impact Ionization MOSFET for Spiking Neural Networks}
\author{Alok Kumar Kamal and Jawar Singh,~\IEEEmembership{ Senior Member,~IEEE}
\thanks{The authors are with the Department
of Electrical Engineering,
Indian Institute Of Technology Patna, Bihar, 801106 India.
E-mail: (kamalalok21@gmail.com; jawar@iitp.ac.in).}}
\maketitle

\begin{abstract}

Silicon bipolar impact ionization MOSFET offers the potential for realization of leaky integrated fire (LIF) neuron due to the presence of parasitic BJT in the floating body. In this work, we have proposed an L shaped gate bipolar impact ionization MOS (L-BIMOS), with reduced breakdown voltage ($V_{B}$ = 1.68 V) and demonstrated the functioning of LIF neuron based on positive feedback mechanism of parasitic BJT. Using 2-D TCAD simulations, we manifest that the proposed L-BIMOS exhibits a low threshold voltage (0.2 V) for firing a spike, and the minimum energy required to fire a single spike for L-BIMOS is calculated to be 0.18 pJ, which makes proposed device $194\times$ more energy efficient than PD-SOI MOSFET silicon neuron (MOSFET silicon neuron) and $5\times10^{3}$ times more energy efficient than analog/digital circuit based conventional neuron. Furthermore, the proposed L-BIMOS silicon neuron exhibits spiking frequency in the GHz range, when the drain is biased at $V_{DG}$ = 2.0 V.

\end{abstract}


\section{Introduction}

The resemblance of the spiking neural network (SNN) with a biological neural network has captivated much attention for the development of neuromorphic computing. The SNN is a third generation artificial neural network (ANN), where the neuron model can convey information by the generated spikes. Moreover, the functionality of spiking neuron is similar to the biological neuron, and one such model is known as the leaky integrated fire (LIF) neuron model. For the implementation of the LIF neuron, conventional analog/digital circuits are widely used~\cite{IG,JV,JW,AJ,PM,RE}. However, such circuit neurons have two significant limitations (a) extensive circuitry (or silicon overhead), and (b) large power consumption.

It is therefore necessary to address these limitations, recently in~\cite{LIFN} the authors have demonstrated a silicon neuron that captures the functioning of the LIF neuron using PD (partially depleted) - SOI (silicon on insulator) MOSFET under impact ionization biasing which utilizes the charge-discharge mechanism of excess carriers in MOSFET silicon neuron. Two mechanisms can enable these excess carriers (a) the impact ionization~\cite{LIFN}, and (b) the band-to-band tunneling (BTBT)~\cite{TC}. The PD-SOI MOSFET silicon neuron is energy efficient in contrast to circuit neuron implementation. However, higher supply (avalanche breakdown) voltage requirements of PD-SOI MOSFET silicon neuron to enable impact ionization (or generate excess carriers) for firing a spike consumes significant power and makes it less energy efficient, also slow floating body charging transients of the MOSFET silicon neuron exhibits slow transients behavior of the order of microseconds~\cite{TK}.

\begin{figure}
		\centering
	\includegraphics[width=90mm,keepaspectratio]{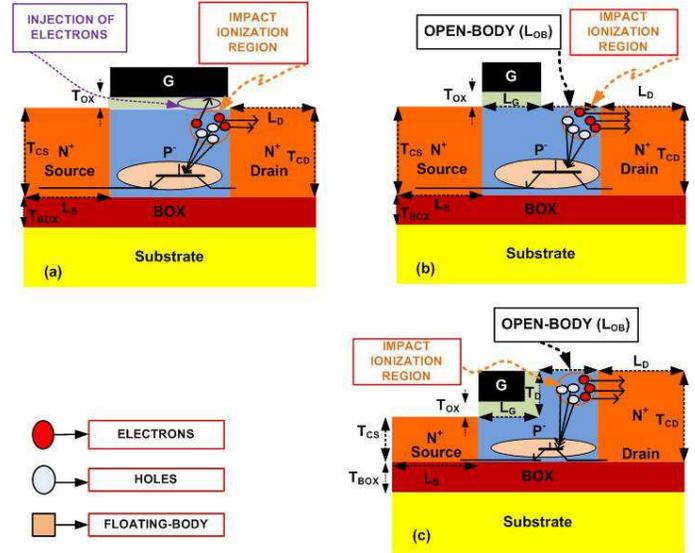}
	\caption{Cross-sectional views of different Si-neurons along with region of impact ionization and floating body: (a) PD-SOI MOSFET silicon neuron~\cite{LIFN}, (b) conventional BIMOS~\cite{MJ}, and (c) proposed L-BIMOS silicon neuron.}
	\label{strfig}
\end{figure}

Therefore, to address the above mentioned limitations of the analog/digital circuit based neurons and the PD-SOI MOSFET silicon neuron, in this paper, first, we investigated a silicon bipolar impact ionization MOS (BIMOS) device which utilizes the positive feedback mechanism of parasitic BJT present in the floating body to realize the ``leaky-integration'' and ``fire'' modes needed for LIF neuron. However, this conventional BIMOS exhibits an avalanche breakdown voltage of $V_{B}$ = 2.85 V~\cite{MJ}, which is extremely high and it may consume huge amount of power to realize a silicon neuron. Thus, we modified the conventional BIMOS structure into `L' shape gate, which results in the crowding of the electric field around the corner of the gate that reduces the avalanche breakdown voltage, referred as L-BIMOS. This `L' shape gate structure of the proposed device also facilitate higher impact generation rate essential for LIF neuron. The proposed L-BIMOS silicon neuron emulates the LIF neuron leaky behavior without any additional circuitry, unlike circuit neurons. It also exhibits significantly low avalanche breakdown voltage in contrast to conventional BIMOS. Thus, the proposed L-BIMOS silicon neuron is compatible with CMOS process, exhibits fast transient behavior, and $194\times$ more energy efficiency as compared to its counterpart PD-SOI MOS silicon neuron.

\section{Device Structures and Simulation Parameters}
Fig.~\ref{strfig} shows the cross-sectional views of (a) PD-SOI MOSFET silicon neuron~\cite{LIFN}, (b) conventional BIMOS, and (c) proposed L gate shaped BIMOS (L-BIMOS) silicon neuron. In the proposed L-BIMOS, the portion of the channel region which is remained uncovered with the gate towards drain terminal is referred as open-body ($L_{OB}$) region, as shown in Fig.~\ref{strfig} (c), whereas in PD-SOI MOSFET silicon neuron whole channel region is covered with gate, as shown Fig.~\ref{strfig} (a). This open-body region facilitate the maximum applied potential $V_{DG}$ across open-body region that results in a high electric field~\cite{KG}. The higher electric filed in this region further enhances the impact ionization rate (IIR) in the proposed L-BIMOS. The enhanced IIR increases the rate of accumulation of holes in the floating body that leads to turn on the parasitic BJT action, hence, fast (order of nanoseconds) transient behavior of LIF neuron. In addition, the generated electrons due to high electric field flow towards drain region without passing under gate oxide, which results in less carrier injection in gate oxide as compared to PD-SOI MOSFET neuron, as shown in Fig.~\ref{strfig}. Further, the gate length of the L-BIMOS is scaled down to 45 nm by keeping $L_{G}$/$L_{OB}$ ratio = 1. We have considered the silicon film thickness underneath the gate and the source region is 25 nm. Remaining simulation parameters and dimensions for both devices are tabulated in Table-I.

\begin{table}
	\centering
	\caption{PARAMETER USED FOR DEVICE SIMULATION}
	\begin{small}

	{\begin{tabular}{|l|c|c|}
	\hline
	\hline
				Parameters & BIMOS & L-BIMOS \\
                \hline
				\hline
				Source doping ($N_{D}$) ($cm^{-3}$) & $10^{20}$ & $10^{20}$ \\
                \hline
                Drain doping ($N_{D}$) ($cm^{-3}$) & $10^{20}$ & $10^{20}$ \\
                \hline
                Channel doping ($N_{A}$) ($cm^{-3}$) & $5\times$$10^{17}$ & $5\times$$10^{17}$ \\
                \hline
                Gate Work function (eV) & 4.6 & 4.6 \\
                \hline
				Gate length  ($L_{G}$) ( nm ) & 75 & 45 \\
                \hline
				Source length ($L_{S}$) ( nm ) & 100 & 100 \\
                \hline
                Drain length ($L_{D}$) ( nm ) & 100 & 100 \\
                \hline
                Open body length ($L_{OB}$) ( nm ) & 75 & 45 \\
                \hline
                Channel Length \\ ($L_{CH}$ = $L_{G}$$+$$L_{OB}$) ( nm ) & 150 & 90 \\
                \hline
                Trench depth ($T_{D}$) ( nm ) & - & 25 \\
                \hline
                Open body depth ($T_{OB}$) ( nm ) & - & 25 \\
                \hline
                Channel thickness at drain side \\ ($T_{CD}$) ( nm ) & 50 & 50 \\
                \hline
                Channel thickness at source side \\ ($T_{CS}$) ( nm ) & 50 & 25 \\
                \hline
                Box thickness ($T_{BOX}$) ( nm ) & 100 & 100 \\
                \hline
                EOT ($T_{OX}$) ( nm ) & 3 & 3 \\
				\hline
				\hline
		\end{tabular}}{}
	\end{small}
\end{table}

The Atlas Silvaco version 3.10.18R is used to simulate the behavior of the proposed device. The device is designed in 2-D background after including parallel electric-field-dependent mobility model, concentration dependent Shockley-Read-Hall model, Fermi-Dirac carrier statistic, standard band to band tunneling (BTBT) model and Masetti mobility models. To initiate impact ionization in the proposed device SELB model is incorporated. In addition, to simulate breakdown phenomenon we have used CURVETRACE algorithm~\cite{AL1}.

\section{Parasitic BJT action in L-BIMOS }

\begin{figure}
\centering
\includegraphics[width=45mm,keepaspectratio]{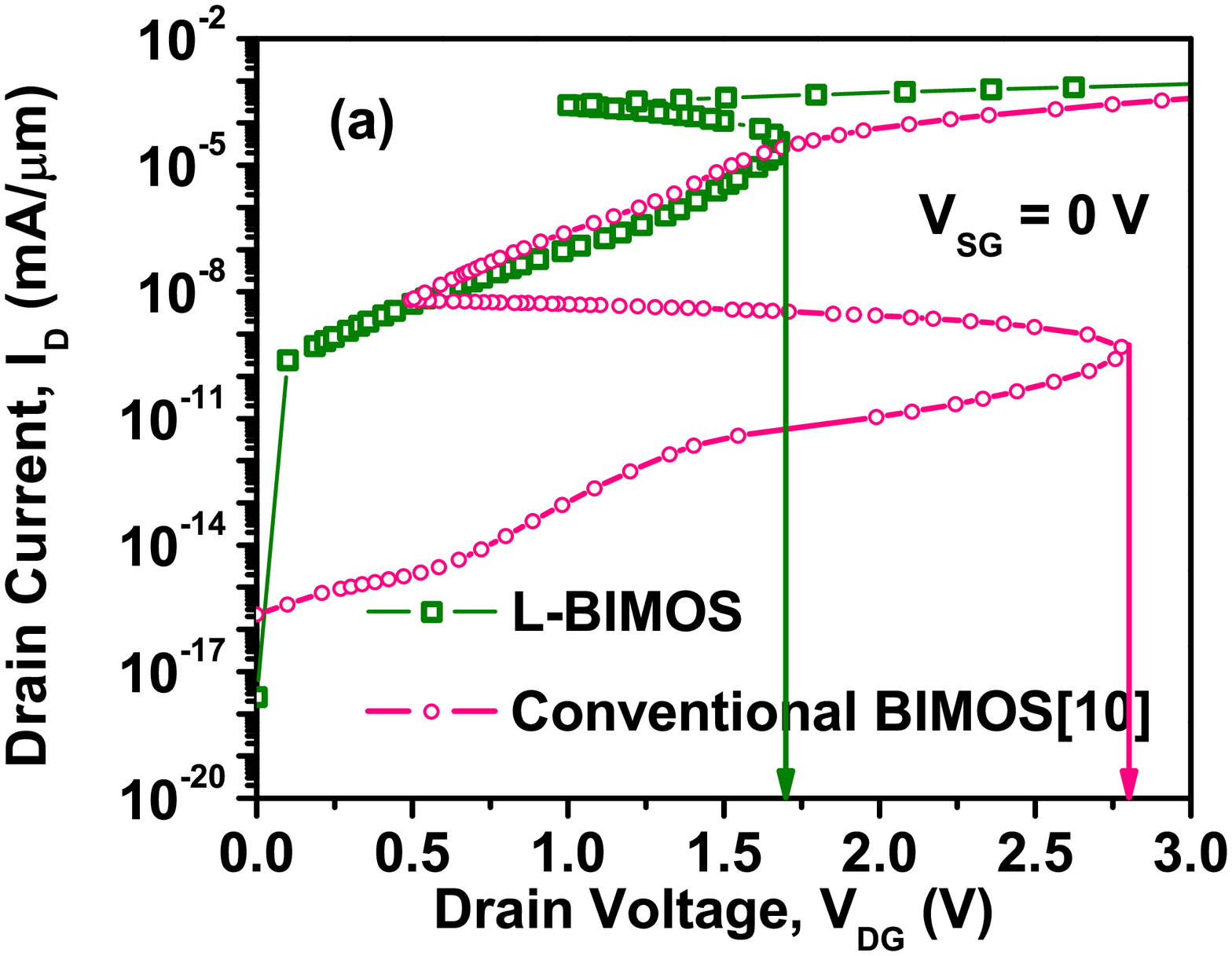}
\includegraphics[width=50mm,keepaspectratio]{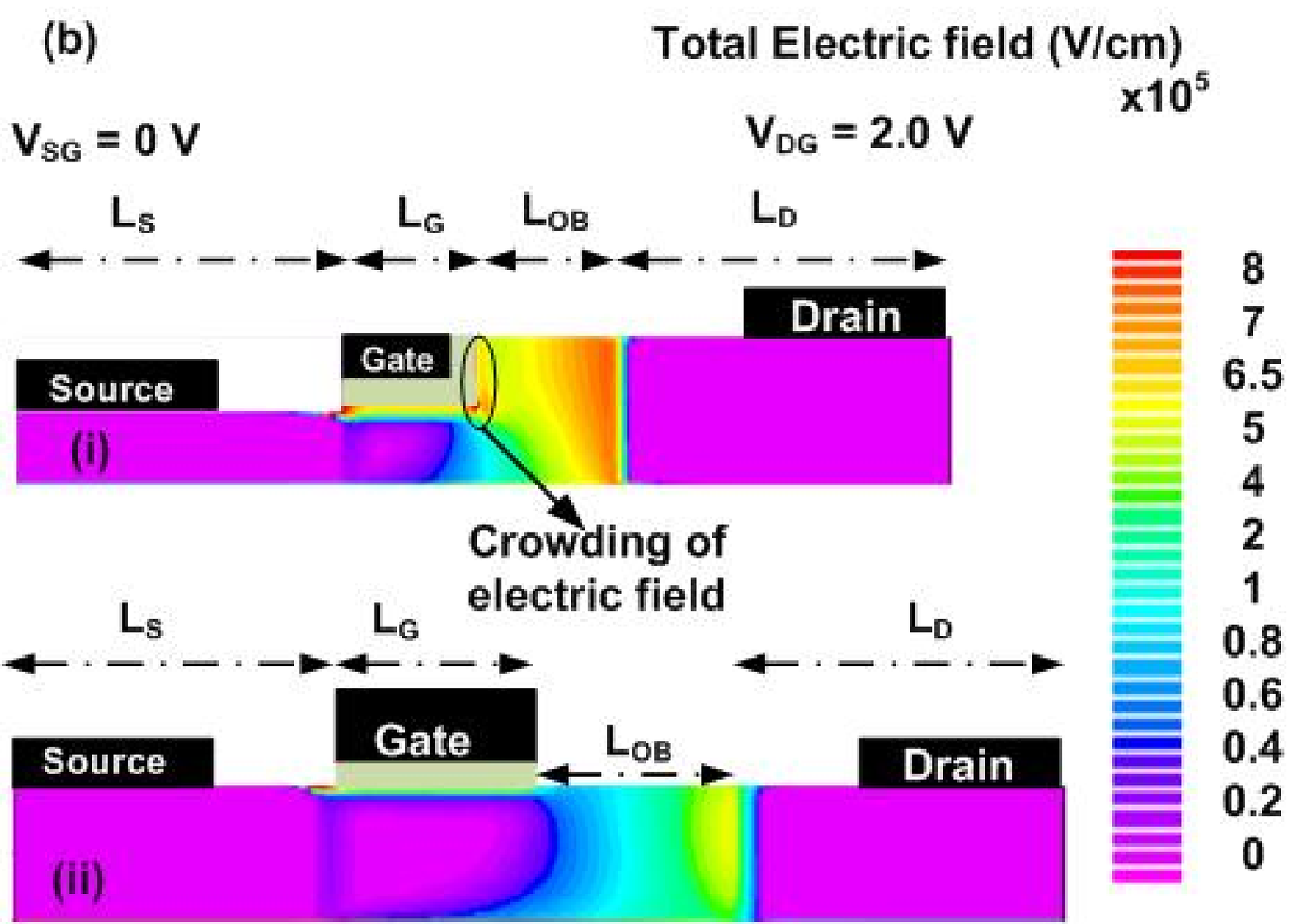}
\caption{(a) Output characteristics comparison of the proposed device (L-BIMOS) with the conventional BIMOS at $V_{SG}$ = 0 V, and (b) electric field comparison between L-BIMOS and conventional BIMOS. }
\label{COMP}
\end{figure}

The snapback observed in the output characteristic confirms the parasitic BJT action contributing to breakdown~\cite{YT}, as shown in Fig.~\ref{COMP}(a). We can see that at $V_{SG}$ = 0 V the breakdown voltage of conventional BIMOS is $V_{B}$ = 2.85 V. The essential parasitic BJT action that describes the charge integration and leaky functionality required for spiking neural network (SNN) to be mimicked artificially by the silicon neuron. However, to recognize conventional BIMOS as a LIF neuron, we have to supply $V_{DG}$ $>$ 2.85 V which is high enough and consumes large amount of power. For the energy efficient use of conventional BIMOS as an LIF neuron, supply voltage should be kept as low as possible, therefore, the breakdown voltage of the conventional BIMOS needs to be reduced further.

To attain the lower value of breakdown voltage, we have proposed a BIMOS by transforming the shape of gate to L shape and named it as L-BIMOS. The proposed device (L-BIMOS) exhibits crowding of electric field at the corner as well as at the edges of gate, and hence, increases electric field at the open-body ($L_{OB}$) region when compared to the conventional BIMOS, as shown in Fig.~\ref{COMP}(b). The increased electric field in open-body ($L_{OB}$) region of L-BIMOS lowers the breakdown voltage to $V_B$ = 1.68 V. Moreover, from Fig.~\ref{COMP}(b) we can deduce that the electric field density in the open body region ($L_{OB}$) of the L-BIMOS is 1.6 times higher than that of conventional BIMOS. The high electric field in L-BIMOS tends to further increases the impact ionization rate (IIR). The increased IIR results in rise of total drain current ($I_D$), this occurs due to accumulation of generated excess holes in floating body region which turns on the parasitic BJT action~\cite{MJ} followed by forward biasing the source channel junction resulting in increase of the total drain current of L-BIMOS (positive feedback mechanism). On the other hand, the generated electrons due to impact ionization escapes towards the highly doped drain side in the absence of gate oxide above it. Therefore, the injection of electrons in gate oxide is minimized in the conventional BIMOS and the proposed L-BIMOS.

\section{L-BIMOS as a LIF Neuron: Results and Discussion}

To replicate biological neuron artificially, the proposed L-BIMOS silicon neuron exhibits charge integration through positive feedback mechanism, and leaky via leakage current that allows L-BIMOS to function as a silicon neuron for SNN. In biological neuron the effect of $Na^{+}$ channels activation and deactivation is same as positive feedback mechanism in L-BIMOS silicon neuron which occurs due to parasitic BJT action. Another important property of biological neuron is its leaky behavior and in L-BIMOS silicon neuron that is incorporated by hole leakage current ($I_{leaky-hole}$) which is produced at source-channel junction. Further, in biological neuron $K^{+}$ channel activation plays major role in resetting the neuron and to reproduce same reset mechanism in L-BIMOS silicon neuron, we reduce $V_{DG}$ to zero volts. Detailed description and analysis of the proposed L-BIMOS silicon neuron is provide in the following sub-sections.

\begin{figure}
\centering
\includegraphics[width=90mm,keepaspectratio]{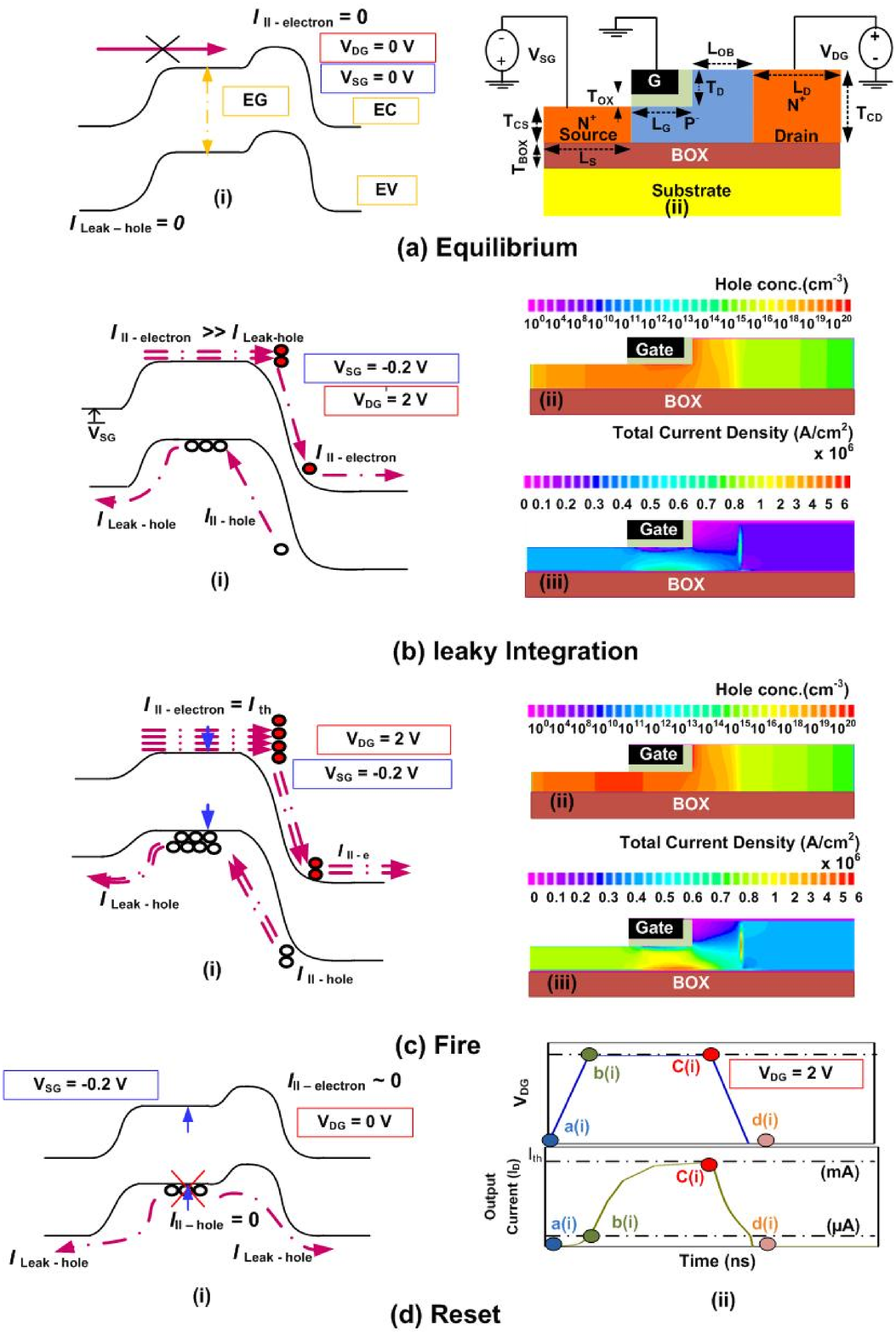}
\caption{(a-i) Equilibrium band diagram and (a-ii) biasing scheme of L-BIMOS. (b-i) Onset of impact ionization (``integration''), results in electron hole pair generation. (b-ii) Hole concentration in channel and (b-iii) total drain current density. (c-i) Barrier lowers at source channel junction with progress of time at fixed bias condition and, the holes start escaping through the source junction (``leaky''). (c-ii) Increase in hole concentration inside the channel and (c-iii) increase in total drain current. (d-i) Energy band diagram at reset condition of L-BIMOS and (d-ii) output spike with respect to input bias.}
\label{BD}
\end{figure}

\subsection{Operation of L-BIMOS silicon neuron}

In this Section, we have discussed the functioning of the proposed device (L-BIMOS) as a LIF neuron with the help of energy band diagrams and carrier concentration, as shown in Fig.~\ref{BD}. The biasing of the proposed device (L-BIMOS) under equilibrium ($V_{SG}$ = $V_{DG}$ = 0 V) and its energy band diagram is shown in Fig.~\ref{BD}(a). From energy band diagram, we can observe that the potential barrier height at the source channel region is very high, as a results there is no thermionic emission of the electron from source side to the drain side. Therefore, from Fig.~\ref{BD}(a-i) we can infer that under this condition the proposed device does not conduct. So in order to start conduction, the impact ionization is initiated with the help of the biasing scheme as shown in Fig.~\ref{BD}(a-ii). In accordance to the biasing scheme the L-BIMOS is biased at $V_{SG}$ = - 0.2 V and $V_{DG}$ = 2.0 V due to which the potential barrier at the source channel region reduces, as shown in Fig.~\ref{BD}(b-i). As a result, the thermionic emission of the electron starts from the source side to drain side of the device. In addition, when these electrons move towards the open-body region, the electrons gets heated up by the applied field. Some of them acquires enough energy to produce impact on covalent bond, thereby new electrons and holes are created. The generated electron moves towards drain, and holes are swept into floating body which results in the electrostatic lowering of the potential barrier at the source-channel junction and due to which the hole leakage current ($I_{leaky-hole}$) starts flowing from the channel region to the source region that defines the ``leaky'' behavior of the proposed L-BIMOS silicon neuron~\cite{LIFN}.

As electron current is greater than the hole leakage current ($I_{II-electron}$ $>$$>$ $I_{leaky-hole}$) shown in Fig.~\ref{BD}(b-i). The hole concentration and the total current density of silicon neuron is shown in Fig.~\ref{BD}(b-ii) and Fig.~\ref{BD}(b-iii). Further, with the progress of time, the source-channel junction shows additional potential barrier lowering as shown in Fig.~\ref{BD}(c-i) due to excess generation of electron and hole pair which results in higher concentration of holes in floating body, as shown in Fig.~\ref{BD}(c-ii). Due to this additional source-channel potential barrier lowering higher rise in total current is observed, as shown in Fig.~\ref{BD}(c-iii) as compared to Fig.~\ref{BD}(b-iii). With the passage of time, further lowering of barrier stops as soon as $I_{leaky-hole}$ becomes equal to $I_{II-hole}$ which leads to total current ($I_D$) saturation in L-BIMOS. Before total current ($I_D$) touches saturation limit, a pre-defined threshold ($I_{th}$ = 0.5 mA/um) is achieved that mimics the ``fire'' mode.

At $I_D$ = $I_{th}$ the proposed device resets by supplying $V_{DG}$ = 0 V. This results in recombination of holes present in channel region and the device is completely at reset condition, as shown in Fig.~\ref{BD}(d-i). In this state, the $I_{leaky-hole}$ and $I_{II-hole}$ currents are zero and the neuron is said to be at reset condition. When the reset is complete the device is again biased with $V_{DG}$ = 2.0 V after certain refractory time to re-initiate the LIF process. Thus, the input voltage $V_{SG}$ produces a response in terms of spike as shows in Fig.~\ref{BD}(d-ii).

\subsection{Impact Ionization in L-BIMOS Silicon Neuron}

\begin{figure}
\centering
\includegraphics[width=45mm,keepaspectratio]{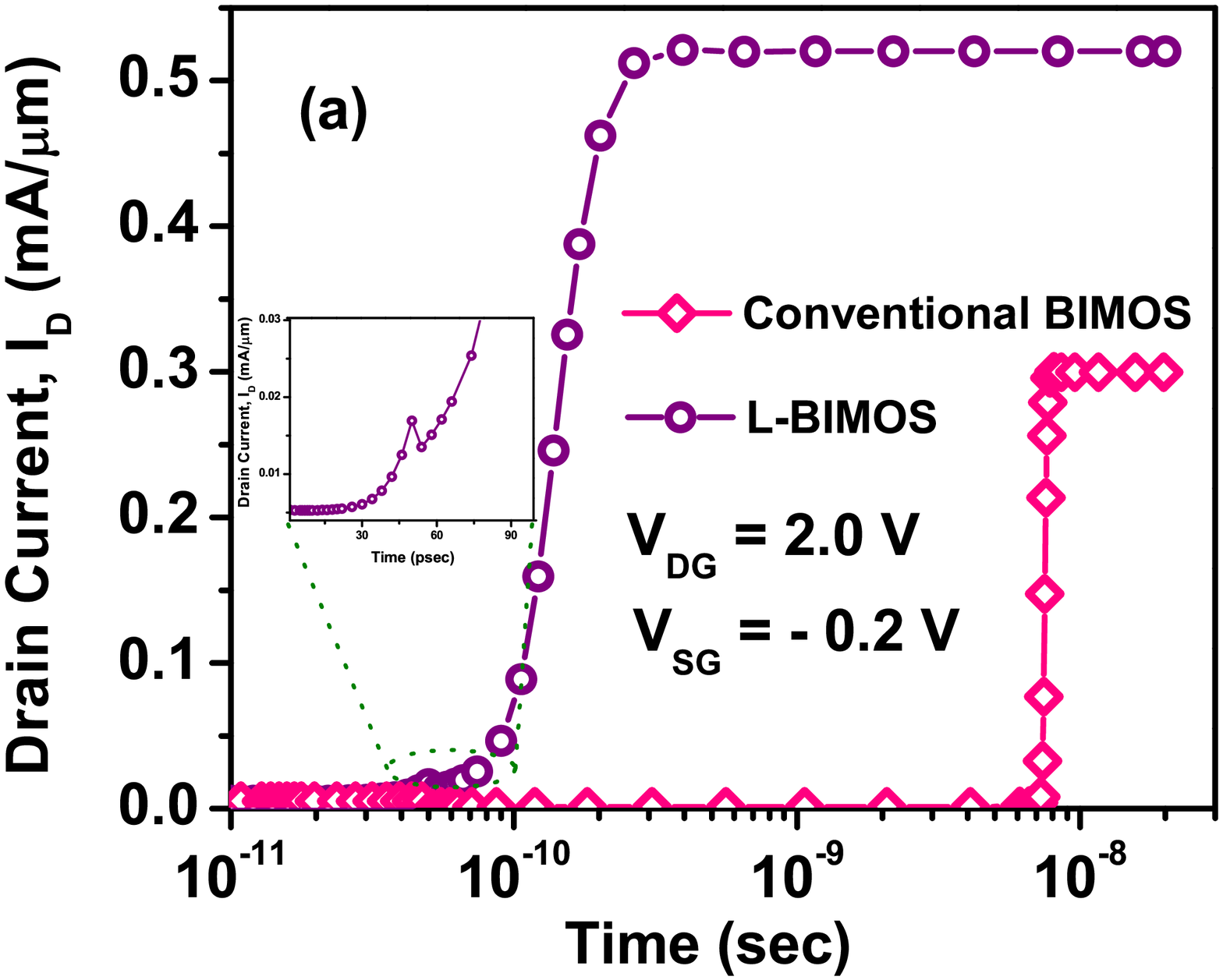}
\includegraphics[width=45mm,keepaspectratio]{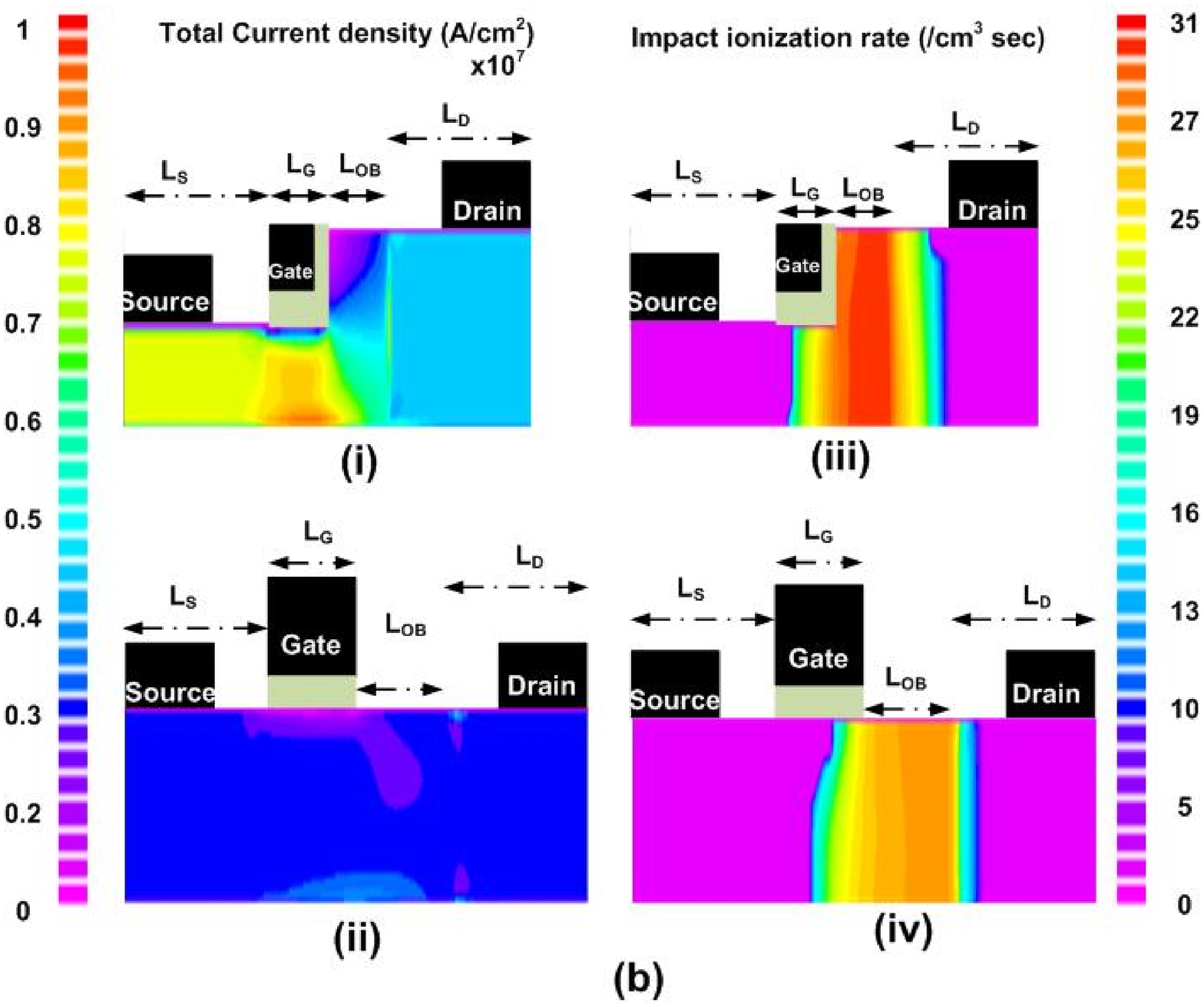}
\caption{(a) Drain current versus transient time characteristics. (b-i) Total current density in L-BIMOS and (b-ii) conventional BIMOS at 1 ns. (b-iii) impact ionization rate in L-BIMOS and (b-iv) in conventional BIMOS at 1 ns.}
\label{idvst}
\end{figure}

The proposed device (L-BIMOS) works on the principle of impact ionization (II) that is responsible for the generation of excess carriers required to work as silicon neuron. The II is a phenomenon in which the electrons from the source side traverses towards the drain side through the channel region. A small amount of these electrons that produces drain current, tends to scatter because of high electric field in open body region, due to which electron-hole pairs generate. The generated electrons moves towards the drain whereas the holes gets accumulated in the floating body resulting in lower potential barrier at source-channel junction.

To initiate the II in L-BIMOS silicon neuron the required bias  ($V_{DG}$ = 2.0 V and $V_{SG}$ = - 0.2 V) is applied. Due to the applied bias there is instantaneous potential barrier lowering which generates the drain current instantly, as shown in inset in Fig.~\ref{idvst} (a). As the potential barrier further lowers due to hole accumulation in floating body that further increases the drain current of L-BIMOS as compared to conventional BIMOS, as shown in Fig.~\ref{idvst}(a) reaches towards the saturation current.

The increase in drain current can be well understood from Fig.~\ref{idvst}(b)(i-ii) in which at 1 ns we can see that the total current density of L-BIMOS is higher as compared to conventional BIMOS because the impact ionization rate (IIR) of L-BIMOS is greater than conventional BIMOS, as shown in Fig.~\ref{idvst}(b)(iii-iv). Hence, due to high IIR in the L-BIMOS, we can observe the fast transient as compared to conventional BIMOS. The fast transients in L-BIMOS results in higher spike frequency as compared to that of conventional BIMOS which makes L-BIMOS more suitable to realize as spiking neural networks (SNN) using silicon neurons.

\begin{figure}
\centering
\includegraphics[width=90mm,keepaspectratio]{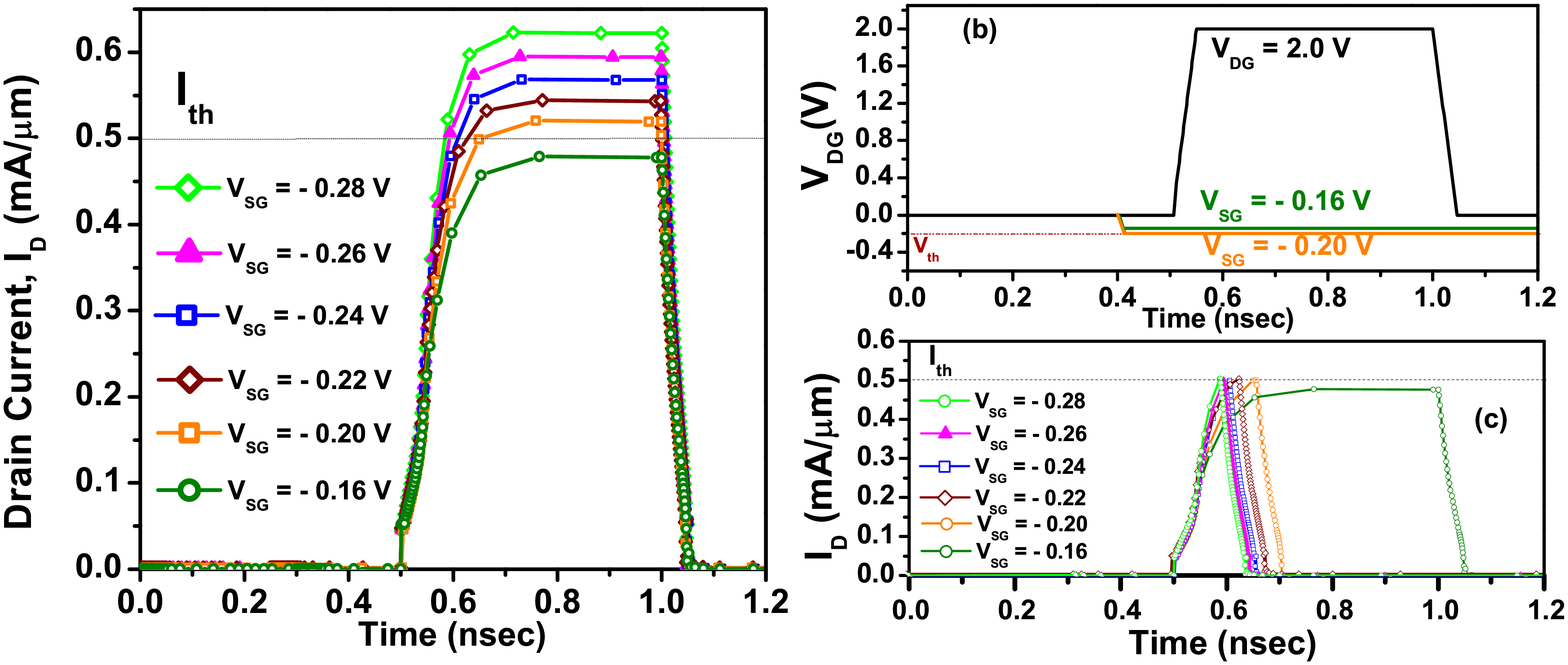}
\caption{(a) Drain Current versus transient time characteristics with constant $V_{DG}$ = 2.0 V. (b-c) For $V_{SG}$ = -0.16 V the $I_{D}$ is unable to spike and it saturates before crossing $I_{th}$ = 0.5 mA/um.}
\label{VDG}
\end{figure}

\subsection{Impact of $V_{SG}$ on Positive Feedback}
For mimicking the neuron behavior, positive feedback plays an important role during integration of drain current over time, under different bias voltages ($V_{SG}$) and $V_{DG}$ is fixed at 2 V is shown in Fig.~\ref{VDG}(a). It can be observed that for certain $V_{SG}$ the $I_D$ reaches a threshold current, $I_{th}$ (fixed at 0.5 mA/um). The $V_{SG}$ corresponding to $I_{th}$ is considered as the threshold voltage ($V_{th}$) of L-BIMOS silicon neuron and it is $\mid V_{th} \mid$ = 0.2 V. As far as $\mid V_{SG} \mid$ is less than $\mid V_{th} \mid$ the positive feedback will not be establish (Fig.~\ref{VDG}(b)). Under this condition the proposed L-BIMOS is unable to generate a sufficient amount of excess holes, which further fails to provide sufficient positive feedback and hence, the total drain current saturates (Fig.~\ref{VDG}(c)) before it reaches $I_{th}$ = 0.5 mA/um, thus no spike is generated. Beside this as we increase $\mid V_{SG} \mid$ $\geq$ $\mid V_{th} \mid$ the total drain current of the proposed device reaches very fast to $I_{th}$ = 0.5 mA/um (``leaky integration'') generating a spike as shown in Fig.~\ref{VDG}(c) and this is due to the presence of sufficient positive feedback (i.e. $\mid V_{SG} \mid$ $\geq$ 0.2 V). On the other hand, in biological neurons this positive feedback mechanism is controlled by $Na^{+}$ channels activation and deactivation. Hence, we can observe that the strength and resemblance of positive feedback mechanism in our proposed device with the biological neuron that depends on the input bias voltage ($V_{SG}$).

\subsection{Spiking Behavior of L-BIMOS Silicon Neuron}
\begin{figure}
\includegraphics[width=42mm,keepaspectratio]{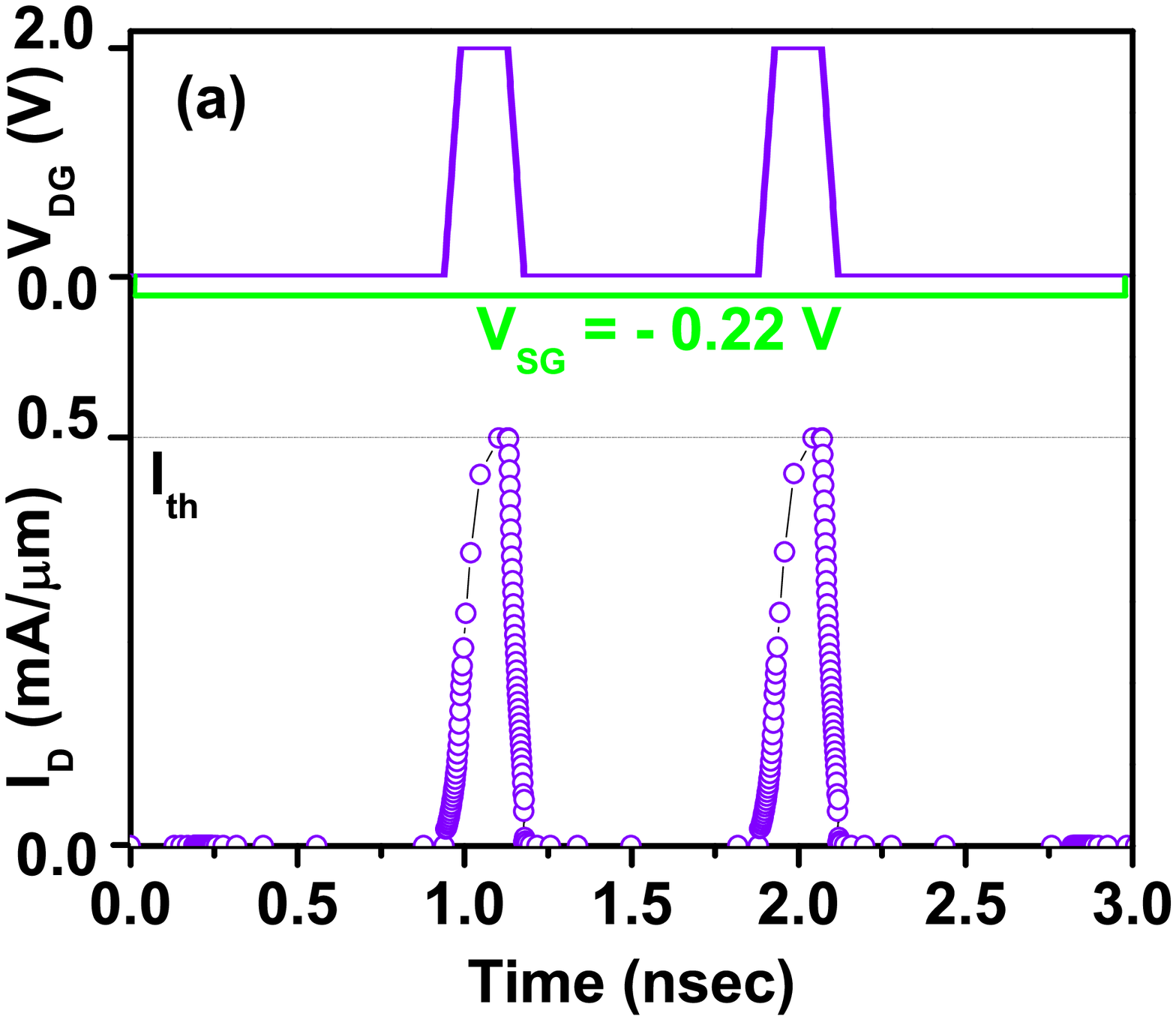}
\includegraphics[width=42mm,keepaspectratio]{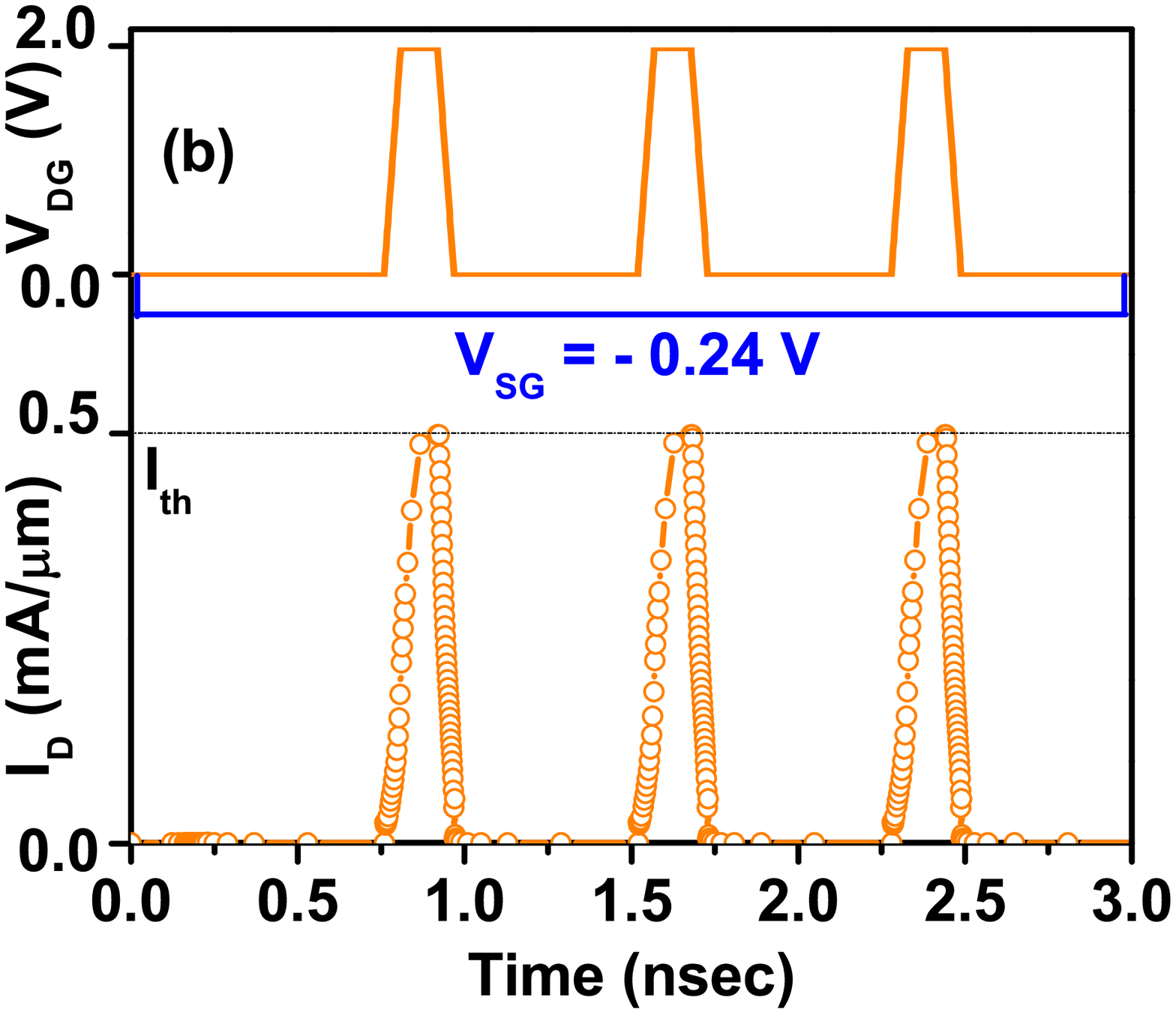}
\includegraphics[width=42mm,keepaspectratio]{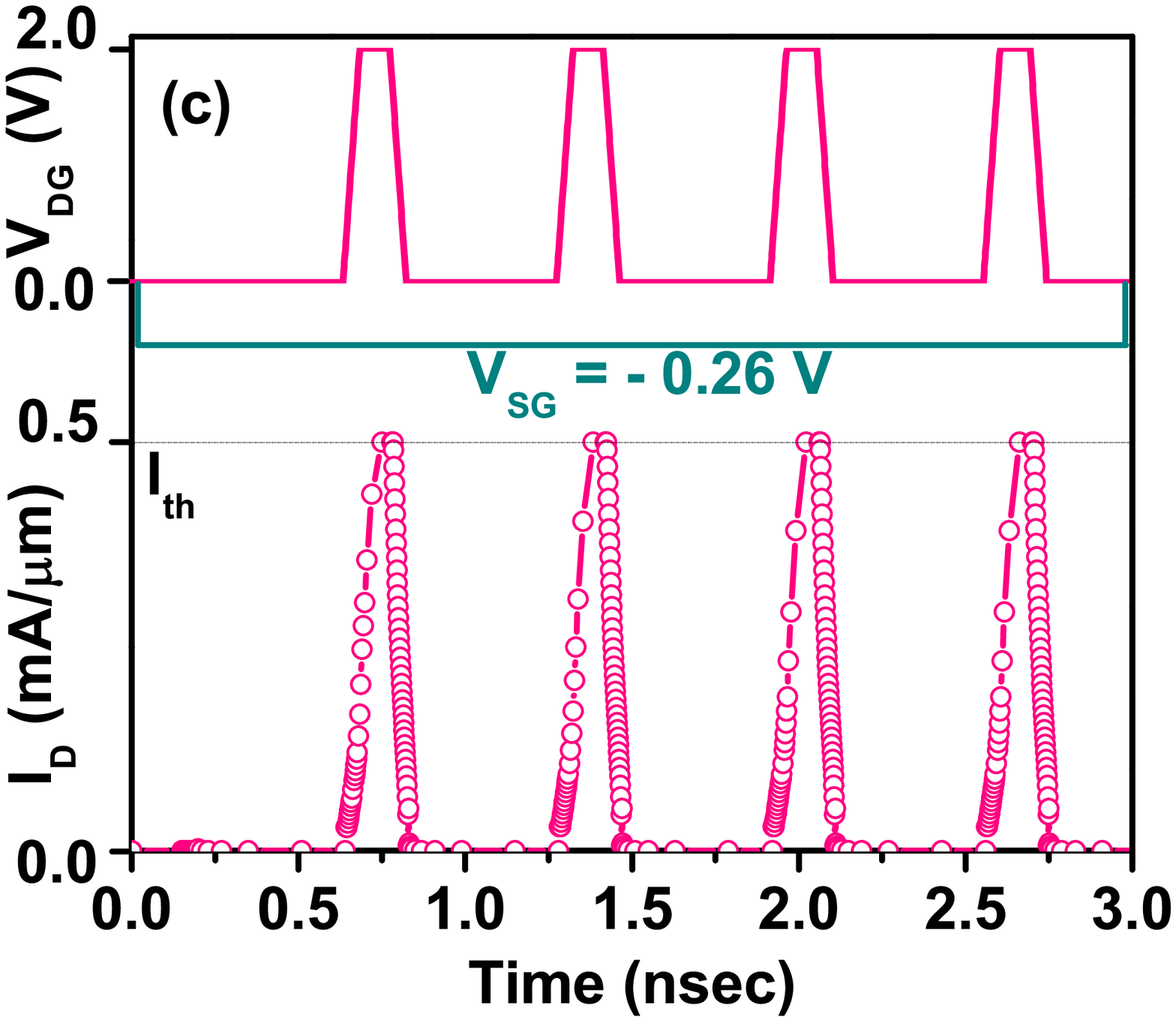}
\includegraphics[width=42mm,keepaspectratio]{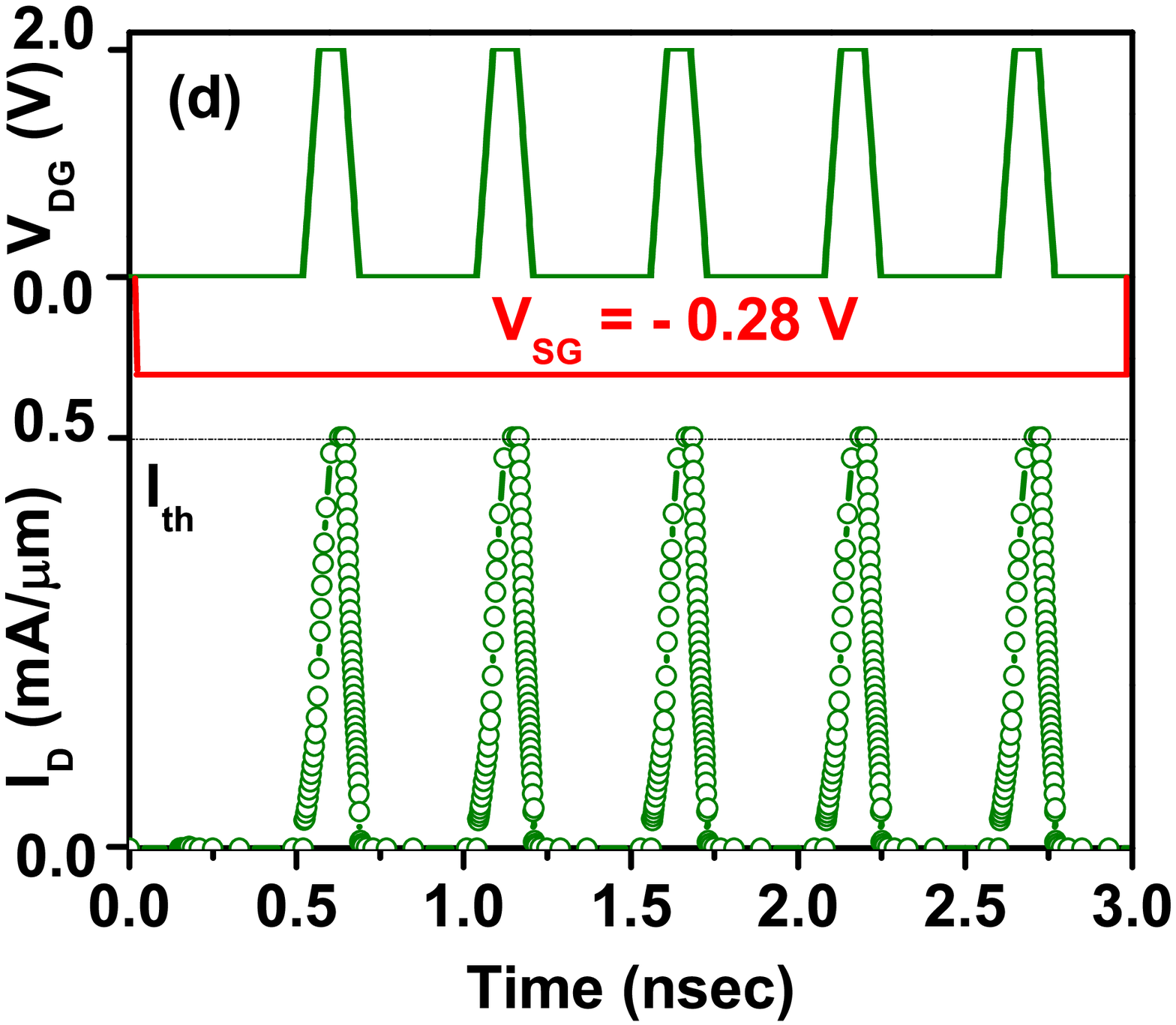}
\caption{Train of spikes at the output of L-BIMOS Si neuron corresponding to different value of $V_{SG}$ at $V_{DG}$ = 2.0 V. }
	\label{spike}
\end{figure}

\begin{figure}
	\centering
	\includegraphics[width=45mm,keepaspectratio]{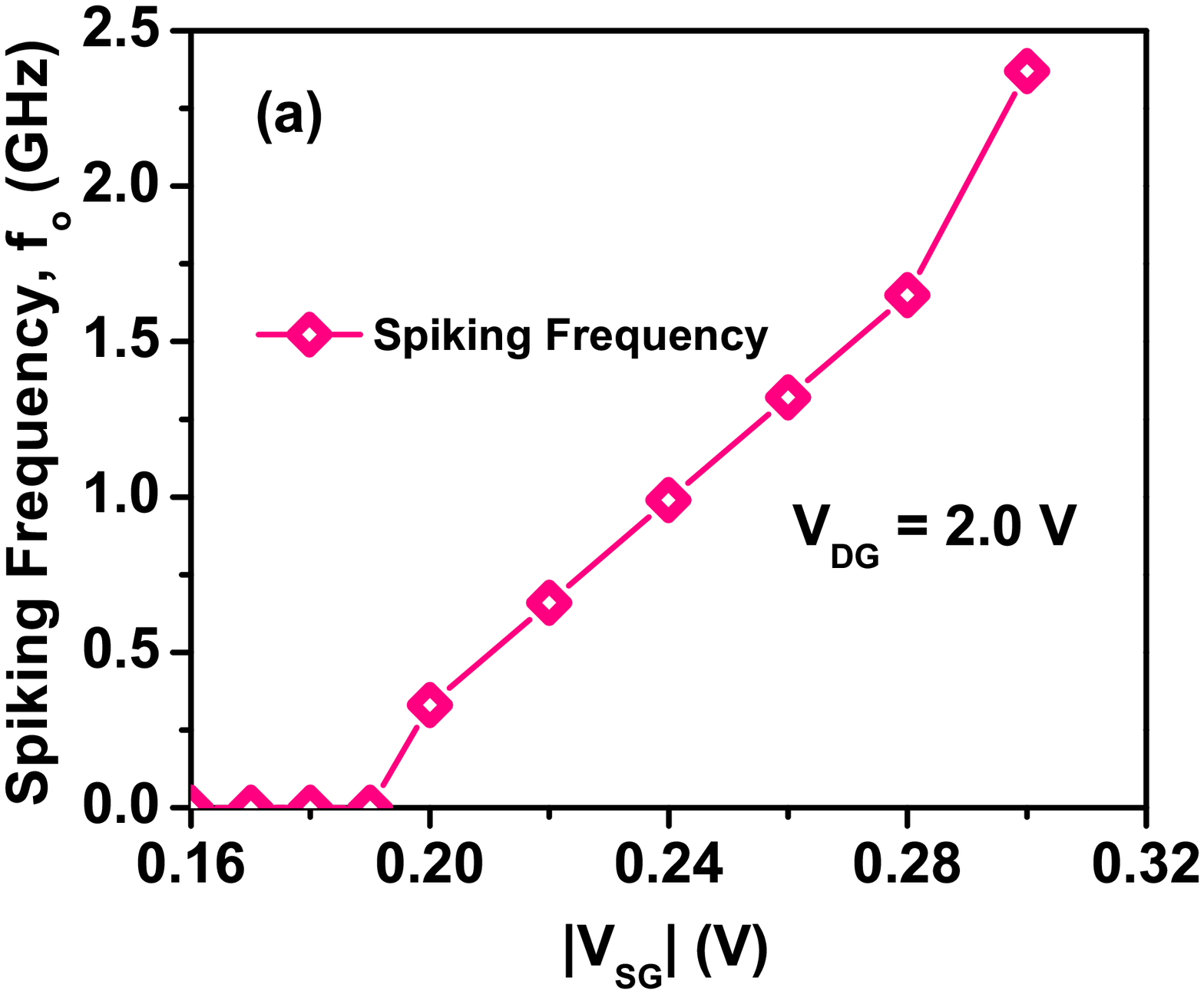}
	\includegraphics[width=45mm,keepaspectratio]{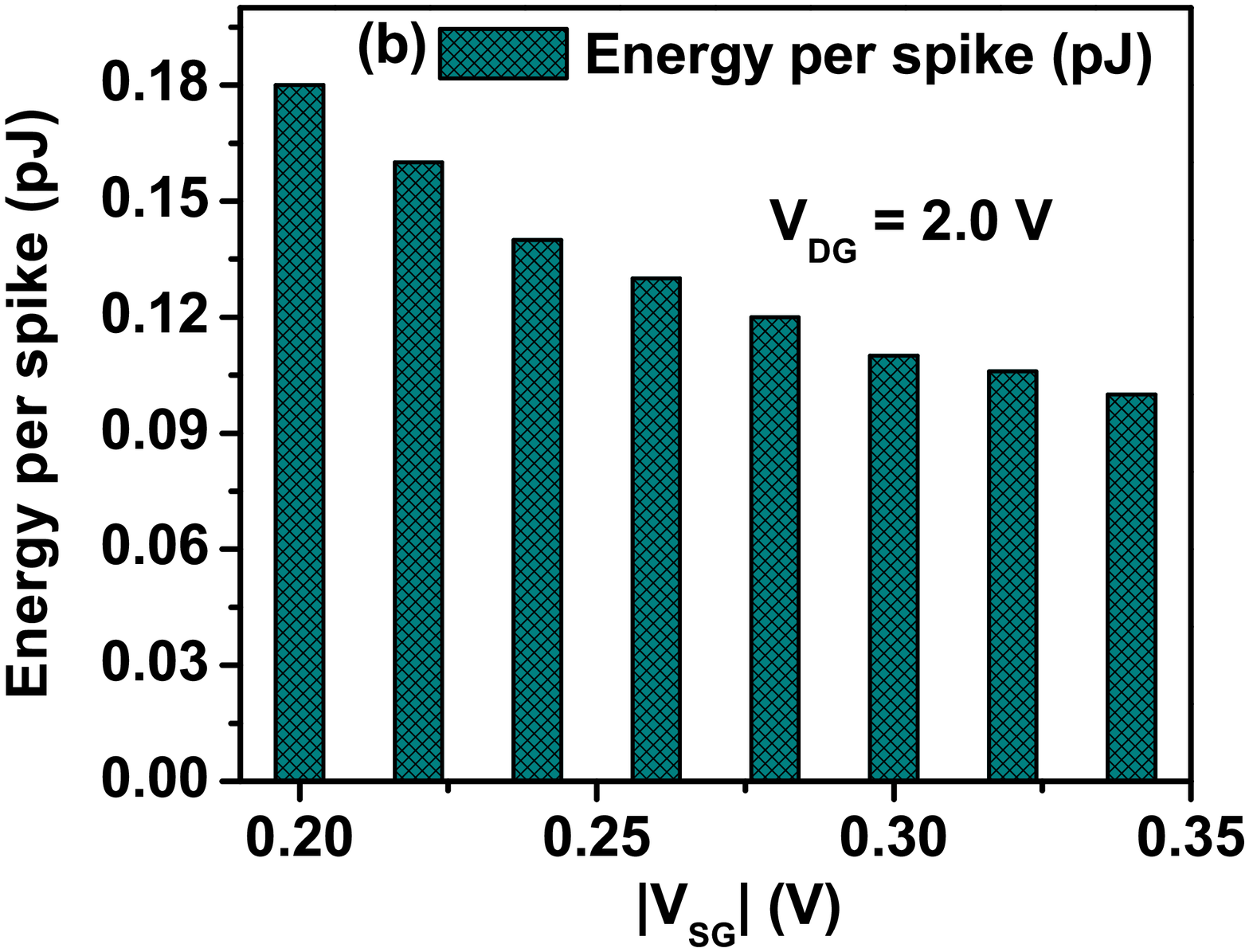}
	\caption{(a) Linear dependence of spiking frequency ($f_{o}$) with input voltage $\mid V_{SG} \mid$ of L-BIMOS silicon neuron and (b) average energy per spike as a function of input voltage $\mid V_{SG} \mid$. }
	\label{freq}
\end{figure}
For realization of spiking neural network, spiking behavior of the proposed device under different biasing conditions ($V_{SG}$ and $V_{DG}$) is shown in Fig.~\ref{spike}.
However, to get spikes $\mid V_{SG} \mid$ should be greater than 0.2 V and $V_{DG}$ must be above breakdown voltage (i.e.1.68 V). For $V_{SG}$ = - 0.22 V and $V_{DG}$ = 2.0 V, the drain current attains $I_{th}$ and as it exceeds $I_{th}$, a spike is fired followed by forced reset i.e. $V_{DG}$ is grounded for 700 ps to flush the stored holes from the floating-body. Now again $V_{DG}$ is set back to 2 V for 140 ps and by doing this the current again start to integrate and fires an identical spike, as shown in Fig.~\ref{spike}(a). As a result, train of spikes can be generated after every 840 ps that dictates the spiking frequency. The number of spikes increases as we increase the $\mid V_{SG} \mid$, which can be seen in Fig.~\ref{spike}(b-d). This increase in number of spikes is due to faster accumulation of excess holes in floating-body with increases in $\mid V_{SG} \mid$ and corresponding to that the $I_{D}$ reaches $I_{th}$ faster. Hence we can achieve higher frequency of spikes with increase in $\mid V_{SG} \mid$.    
\par
As long as $\mid V_{SG} \mid$ $<$ 0.2 V zero spiking frequency ($f_{o}$) is observed at the output of proposed L-BIMOS silicon neuron, as shown in Fig.~\ref{freq}(a). This is due to the insufficient electron supply from the source region to the drain region which result in zero spikes. However, for $\mid V_{SG} \mid$ $\geq$ 0.2 V, the drain current of the device achieves the threshold current i.e $I_{th}$ = 0.5 mA/um, results generation of spikes. As we increases the $\mid V_{SG} \mid$ beyond 0.2 V, the $f_{o}$ increases monotonically, as shown in Fig.~\ref{freq}(a). Thus, proposed device offers the signature of a biological neuron, i.e., for $\mid V_{SG} \mid$ $<$ 0.2 V the device gives zero $f_{o}$ whereas for $\mid V_{SG} \mid$ $\geq$ 0.2 V $f_{o}$ increases linearly.

\par
We have also compared the performance of the proposed device with the previous published CMOS process compatible works, as mentioned in Table-II. The CMOS circuit based neuron~\cite{JW,AJ} consist of leaky integrate circuit which increases the complexity of SNN as well as the energy consumption per spikes. So in this work we have replaced the leaky integration function of neuron with the proposed L-BIMOS silicon neuron. One can see that the proposed L-BIMOS silicon neuron requires 0.2 V for firing a spike, which is 60 mV less when compared to previous reported LIF neuron~\cite{LIFN} and hence makes the proposed LIF neuron energy efficient. For estimating the energy per spike, we have adopted the following expression~\cite{BD}:

\begin{equation}
E_{spike} = V_{spike} \times t_{spike} \times I_{th}
\end{equation}

The maximum energy/spike for the proposed L-BIMOS silicon neuron biased at $V_{DG}$ = 2.0 V and $V_{SG}$ = - 0.2 V is calculated to be 0.18 which is $194\times$ less than the PD-SOI MOS silicon neuron. In addition we have observed that with the increase in $\mid V_{SG} \mid$ the average energy per spike decreases as shown in Fig.~\ref{freq}(b). This decrease in energy/spikes is due to the fast leaky integration of $I_{D}$ with increase in $\mid V_{SG} \mid$.
\begin{table}
	\centering
	\caption{BENCHMARK COMPARISON OF LIF NEURON}
	\begin{small}
	{\begin{tabular}{|l|c|c|c|c|c|}
	\hline
	\hline

				References & Device Type & $L_{CH}$ & $\mid V_{th} \mid$ & Energy/spike  \\
                \hline
				\hline
				 ~\cite{IG} & CMOS & - & $>$1 V & 900 pJ \\
                 ~\cite{JW} & CMOS & - & - & 8.5-9 pJ \\
                 ~\cite{AJ} & CMOS & - & - & 41.3 pJ \\
                 ~\cite{TT}& PCM & - & - & 5 pJ\\
                 ~\cite{LIFN}& SOI MOS & 100 nm & 0.26 V & 35 pJ \\
                 ~\cite{SL}& PCMO RRAM & - & - & 4.8 pJ \\
                This Work & L-BIMOS & 90 nm & 0.2 V & 0.18 pJ \\
				\hline
				\hline
		\end{tabular}}{}
	\end{small}
\end{table}
\section{Conclusion}
In this paper, we have demonstrated the functioning of LIF neuron based on CMOS process compatible L shaped gate bipolar I-MOS (L-BIMOS). The proposed device exhibits reduced breakdown voltage $V_{B}$ = 1.68 V, and consumes 0.18 pJ/spike energy that makes it $194\times$ more efficient in energy in contrast to its counterpart PD-SOI MOS silicon neuron for same functionality. Further, we manifest that the proposed L-BIMOS silicon neuron offers the signature of a biological neuron that can be exploited for the complex spiking neural networks (SNN). The spiking frequency increases linearly with increase in $\mid V_{SG} \mid$ above threshold voltage, and the proposed device exhibits higher spiking frequency of the order of GHz at a lower supply voltage as compared to PD-SOI MOS silicon neuron (MHz). Thus, L-BIMOS silicon neuron has a potential for realization of massive parallel processing required for SNN.

\end{document}